\documentclass[12pt]{iopart}
\newcommand{\pd}[2]{\frac{\partial #1}{\partial #2}}
\newcommand{\gyav}[1]{\left\langle #1 \right\rangle_\mathbf{R}}

\newcommand{\eqref}[1]{(\ref{#1})}

\newcommand{\dv}{\, \mathrm{d}^3 \mathbf{v}}
\newcommand{\turbav}[1]{\left\langle #1 \right\rangle_t}

\newcommand{\jacobv}{\mathcal{J}_v}
\usepackage{color,soul}
\definecolor{red}{rgb}{1.0,0,0}
\setulcolor{red}
\usepackage{iopams}  
\usepackage{multicol}  
\usepackage{graphicx}
\begin{document}

\title[Anomalous transport of fast ions ]{Global anomalous transport of ICRH- and NBI-heated fast ions}

\author{G. J. Wilkie$^{1}$, I. Pusztai$^1$, I. Abel$^2$, W. Dorland$^3$, T. F\"ul\"op$^1$}

\address{$^1$ Department of Physics, Chalmers University of Technology, Gothenburg, Sweden}
\address{$^2$ Princeton Center for Theoretical Science, Princeton, NJ, United States}
\address{$^3$ Department of Physics, University of Maryland, College Park, MD, United States}

\ead{wilkie@chalmers.se}
\vspace{10pt}
\begin{indented}
\item[] Submitted 25 Aug 2016
\end{indented}

\begin{abstract}

  By taking advantage of the trace approximation, one can gain an
  enormous computational advantage when solving for the global
  turbulent transport of impurities. In particular, this makes
  feasible the study of non-Maxwellian transport coupled in radius and
  energy, allowing collisions and transport to be accounted for on
  similar time scales, as occurs for fast ions. In this work, we study
  the fully-nonlinear ITG-driven trace turbulent transport of locally
  heated and injected fast ions. Previous results indicated the
  existence of MeV-range minorities heated by cyclotron resonance, and
  an associated density pinch effect. Here, we build upon this result
  using the \textsc{t3core} code to solve for the distribution of
  these minorities, consistently including the effects of collisions,
  gyrokinetic turbulence, and heating. Using the same tool to study
  the transport of injected fast ions, we contrast the qualitative
  features of their transport with that of the heated minorities.
  Our results indicate that heated minorities are more strongly affected by microturbulence than injected fast ions. The physical interpretation of this difference provides a possible explanation for the observed synergy when NBI heating is combined with ICRH.
  Furthermore, we move beyond the trace approximation to develop a
  model which allows one to easily account for the reduction of
  anomalous transport due to the presence of fast ions in
  electrostatic turbulence.

\end{abstract}

%
%
%
%
%

Fast ions are an important component of a fusion device, being
responsible for a portion of heating required to bring tokamaks and
stellarators up to fusion-relevant temperatures. Various phenomena can
cause radial transport and hence a redistribution of this energy,
including: neoclassical transport, transport from nonaxisymmetric
magnetic ``ripples'', stochastic magnetic field regions, Alfv\'en
waves driven unstable by the fast ions themselves, and
microturbulence. This latter effect is what we focus on in this work,
taking advantage of recently developed tools to study the coupled
radius-energy phase space transport of trace non-Maxwellian species.

With the ion cyclotron resonance heating (ICRH) technique,
electromagnetic waves are launched into the plasma at a frequency
resonant with that of ion cyclotron motion at some locations. Under
certain circumstances, a very small population of minority ions can
very efficiently absorb the power and be heated up to MeV-range
energies \cite{kazakov_resonant_2015}. Previous results indicated that
the effect of microturbulence on these heated minorities is to cause a
density ``pinch'' of fast ions against their temperature gradient
\cite{pusztai_turbulent_2016}, an effect which is also expected from
the quasilinear diffusion coefficients of Maxwellian fast ions
\cite{angioni_gyrokinetic_2008}. Another auxiliary heating scheme is
neutral beam injection (NBI) in which an energetic beam of neutral
particles is injected into the plasma, where they collide with thermal
ions, ionizing and giving up their energy. This effectively results in
a high energy ``source'' of fast ions taken from the bulk thermal
population, mimicking the birth of alpha particles in fusion
reactions.

We consider two classes of fast ions in this work: ``heated'', where
heat is supplied to an already-present species; and ``injected'',
which involves fast ions entering the plasma at high energy and
slowing-down via collisions. The ICRH-heated minorities are an example
of the former category, while NBI ions and alpha particles are
examples of the latter. Although NBI heating profiles are unlikely
to be similar to ICRH ions in real experiments, to assist in directly
comparing both types, we will be using the same boundary conditions
and energy deposition profile. 

This article is organized as follows. In Sec.~1, we 
introduce our framework for solving the transport problem and describe
the case we simulate. In Sec.~2, we describe and interpret our results
from the \textsc{t3core} transport simulations, contrasting the
behavior of heated and injected fast ions. Finally, in Sec.~3, we
improve upon the trace approximation by developing a simple model to
adjust the bulk plasma anomalous transport when fast ions are present
in significant quantities.

\section{Description of the problem}

The method we use to solve the multiscale transport problem is based
on the trace approximation, in which the fast ions do not have an
effect on the turbulence. In a later section, we provide corrections
to this approximation. For now, assuming the trace approximation
holds, the pitch-angle averaged low-collisionality transport equation
reads \cite{wilkie_microturbulent_2015,waltz_quasilinear_2013}:
\begin{equation} \label{transporteqn}
   \pd{F_0}{t} + \frac{1}{V'} \pd{ }{r} V' \Gamma_r + \frac{1}{v^2} \pd{ }{v} v^2 \Gamma_v = S,
\end{equation}
where $S$ is the source; $V(r)$ is the volume enclosed by the flux
surface labelled by its half-width $r$; $\Gamma_r$ is the turbulent
radial flux of fast ions as a function of speed $v$; and $\Gamma_v$
includes all forms of transport in energy: turbulent energy exchange,
direct heating, collisional diffusion and slowing-down. Upon
pitch-angle averaging to obtain Eq.~\eqref{transporteqn}, we eliminate
the pitch-angle scattering from both collisions and
turbulence. However, in order to obtain a closed equation for the
pitch-angle averaged distribution function $F_0$, we neglect the pitch
angle dependence of the turbulent diffusivities (which multiply
derivatives of $F_0$). According to the scalings of
Ref.~\cite{hauff_electrostatic_2009}, this assumption is
incorrect. However, the numerical results of
Ref.~\cite{pueschel_anomalous_2012} seem to indicate a very weak
dependence of the diffusivity on pitch angle, at least at high energy
and for non-extreme pitch
angles (\emph{e.g.}~$\xi \lesssim 0.99$). With the caveat that a more complete treatment includes the
pitch-angle dependence of the turbulent transport, we proceed to apply
the trace approximation \cite{wilkie_validating_2015} to write the
fluxes in terms of energy-dependent diffusivities as:
\begin{equation} \label{rflux}
   \Gamma_r = - D_{rr} \pd{F_0}{r} - D_{rv} \pd{F_0}{v}
\end{equation}
\begin{equation} \label{vflux}
   \Gamma_v = - D_{vr} \pd{F_0}{r}  - v \nu_s F_0 - \left( \frac{1}{2} v^2 \nu_\| + D_{vv} + \frac{P}{3 n_f m} \right) \pd{F_0}{v},
\end{equation}
where $\nu_s$ and $\nu_\|$ are the slowing-down and parallel velocity
diffusion collision frequencies \cite{helander_collisional_2002}
summed over all bulk species, $P(r)$ is the heating power per unit
volume, $m$ is the fast ion mass, and $n_{f}$ is the local fast ion 
density. Whether the source $S$ or injected power $P$ is used to
inject energy into the system is the difference between ``injected''
and ``heated'' fast ions discussed above, and this is a distinction we
will continue to make throughout this work. The form of the heating
term in Eq.~\eqref{vflux} is the same as the equation used by
Ref.~\cite{stix_fast-wave_1975} for the isotropic part of the
ICRH-heated distribution.  Equations \eqref{transporteqn},
\eqref{rflux}, and \eqref{vflux} form a closed two-dimensional partial
differential equation and is of similar form to that used in
Ref.~\cite{waltz_quasilinear_2013}, except our treatment is not a
quasilinear model, but is rigorous for trace transport in full
nonlinear turbulence. The tool we employ to solve
Eq.~\eqref{transporteqn} is \textsc{t3core}, which was originally
developed for alpha particle transport \cite{wilkie_transport_2016}
and employs a finite-volume method and obtains the diffusion
coefficients by post-processing output from \textsc{gs2}
\cite{dorland_electron_2000} nonlinear gyrokinetic turbulence
simulations. This is done by comparing the fluxes of two different
trace species with the same charge and mass, but different radial gradients
and/or temperatures. For each energy and radius, one can determine the
unknowns $D_{rr}$ and $D_{rv}$ by algebraically solving
Eq.~\eqref{rflux} for each of the two species given the two different
equilibrium distributions and the corresponding calculated fluxes.

The local properties of our nominal case at $r = 0.25a$, which is the same ITER-like scenario from Ref.~\cite{pusztai_turbulent_2016}, are as 
follows. The shape of the flux surface is such that the ellipticity is
$\kappa = 1.409$ and triangularity is $\delta = 0.075$ (with radial
derivatives $a \kappa' = 0.0914$ and $a \delta'=0.1405$,
respectively). The center of the flux surface has a major radius of $R
= 3.29 a$, and $R'(r) = -0.0818$. The magnetic shear is $\hat{s} =
0.127$ and $q = 1.27$. The plasma is a mixture of 70\% protons and
30\% $^4$He by charge density, there is no local gradient in particle
density for either electrons or ions. Electrons are a dynamic kinetic
species, and our simulations include only electrostatic
fluctuations. This case is based off of an ITER hybrid scenario (case
20020100 from the ITPA public database \cite{roach_2008_2008}), but 
with an increased ion temperature gradient scale length $a T_i' / T_i
= -1.509$ so that the ion temperature gradient (ITG) mode is
marginally unstable. The electron density is $9.24\times 10^{19} /
\mathrm{m}^3$, temperatures are $T_i = T_e = 25 \,\mathrm{keV}$, and
$\rho^* = \rho_i / a = 0.0033$.

In our transport simulations for the fast ions, an estimated peak power density of $P_\mathrm{max} = 140 \,\mathrm{kW}/\mathrm{m}^3$ is
absorbed by the $^3$He minority, and is radially distributed as a
Gaussian with a width of $\Delta r = 0.025 a$: $P(r) = P_\mathrm{max} \exp\left[ - \left( r - 0.25 a \right)^2 / 2 \Delta r^2 \right]$. This is an
approximation of the simulated absorption profile in
Ref.~\cite{pusztai_turbulent_2016}. When we consider the NBI-like
injected case, these are protons injected at 1 MeV with an identical
radial power distribution as the ICRH ions:  $S\left( r,v \right) \propto \exp\left[ - m_f \left(v - v_\mathrm{inj} \right)^2 / 2 T_i \right]$, where the constant of proportionality for each radius is numerically calculated so that $\int S(r,v) \left(m_f v^2 /2 \right) \,\mathrm{d}^3\mathbf{v} = P(r)$.

\begin{figure}
\centering
   \includegraphics[width=0.5\textwidth]{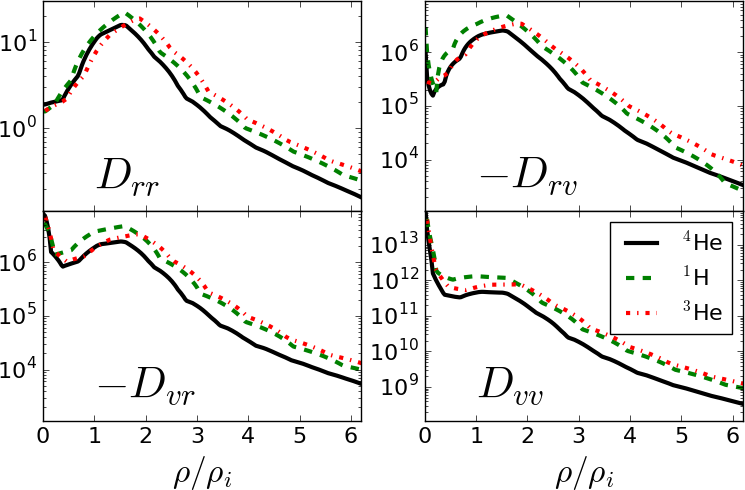}
   \caption{\label{diffcoeff} The diffusion coefficients of Eqs.~\eqref{rflux} and \eqref{vflux} as functions of energy, for which the Larmor radius $\rho$ is used a proxy. Determined from
     nonlinear \textsc{gs2} simulations for the three types of trace fast
     ions considered in this work: alpha particles (black solid), NBI
     knock-on hydrogen (green dashed), and ICRH-heated $^3$He
     minorities (red dotted). Units for $D_{rr}$ are
     $\mathrm{m}^2/\mathrm{s}$, $D_{rv}$ and $D_{vr}$ are given in
     $\mathrm{m}^2/\mathrm{s}^2$, and $D_{vv}$ in
     $\mathrm{m^2}/\mathrm{s}^3$. }
\end{figure}

The domain of our transport simulations is the annulus
$0.1 \leq r/a \leq 0.4$. Throughout this domain, the turbulence has identical properties as calculated by gyrokinetic flux tube
simulations at $r = 0.25a$. This is done in order to isolate the
effects of transport; although a transport simulation whose flux tubes are radially-varying 
is more reflective of reality, and certainly possible in our framework, it is an unecessary complication. Figure \ref{diffcoeff} shows the energy dependence of
the four diffusion coefficients as determined from these \textsc{gs2}
simulations. There, we compare several types of fast ions which might
be present: ICRH-heated $^3$He, NBI knock-on $^1$H, and fusion
produced alpha particles ($^4$He).  The horizontal axis is expressed
as the ratio of the Larmor radius (at a given energy) to the thermal
hydrogen Larmor radius $\rho_i$. This is done to highlight the
consistent physics between the species: the diffusion peaks at
approximately the same Larmor radius in all cases, and has similar
scaling at high energy. Although the coefficients differ by factors of
order unity, qualitatively they are similar functions of energy, when
properly scaled to account for different isotopes. These coefficients
form the basis of the transport simulations that follow. Note that
despite $D_{vr}$ and $D_{vv}$ appearing large when written in SI
units, they contribute negligibly to the transport simulations because these terms are small compared to the corresponding radial transport terms and collisions. Nevertheless, these terms are 
retained in our simulations for completeness. For a disussion of Onsager symmetry between $D_{rv}$ and $D_{vr}$ at high energy, see the appendix.

From our gyrokinetic simulations of this case, we obtain a turbulent
heat flux of $0.73 \mathrm{MW}/\mathrm{m}^2$, which translates to
about 100 MW of core heating power, which is beyond ITER's external
heating capability. This indicates that our simulations are too
strongly driven. Nevertheless, we will consider this the ``nominal''
case, and we will routinely rescale the diffusion coefficients by
various factors throughout this work to quantitatively examine the
range of effect turbulence can have on fast ions. The energy
dependence of the diffusion coefficients does not change significantly
when more strongly- or weakly-driven turbulence simulations are run,
so this practice of rescaling the diffusion coefficients is robust.

The boundary conditions used for the transport solution are as
follows. At $r = 0.4 a$, the population of minorities is held fixed as
a Maxwellian at the local ion temperature and a density of
$n_{f0} = 10^{17}/\mathrm{m}^3$, approximately 0.1\% of the electron density
$n_e = 9.24\times10^{19}/\mathrm{m}^3$. At the inner boundary
$r = 0.1a$, the net flux of minorities into the domain is zero at
every energy. In velocity space, the flux through $v=0$ vanishes, and the distribution is set to vanish at a suitably high speed $v_\mathrm{max}$. The temperature is held fixed at
$T_e = T_i = 25 \mathrm{keV}$ throughout the domain.

\section{Comparing heated and injected ions: results and
  interpretation}

Previous results indicated that microturbulence has the effect of
causing a particle pinch of ICRH-heated minority ions against their
own strong temperature gradient \cite{pusztai_turbulent_2016}. This conclusion was reached with a fixed fast ion distribution and radial profiles, while here we let these evolve consistently with the turbulence around the peak heating location. For comparison, we will
be comparing this to ions injected at high energy, reflecting the
behavior of neutral beam knock-on ions. Specifically, the velocity distribution of
ICRH-type ions is more ``globally'' affected and moments are generally
more sensitive to the turbulence, whereas for injected fast ions,
turbulence has a localized effect where transport is the strongest,
causing an inversion (``bump-on-tail'') in some cases.

\begin{figure}
\centering
   \includegraphics[width=0.5\textwidth]{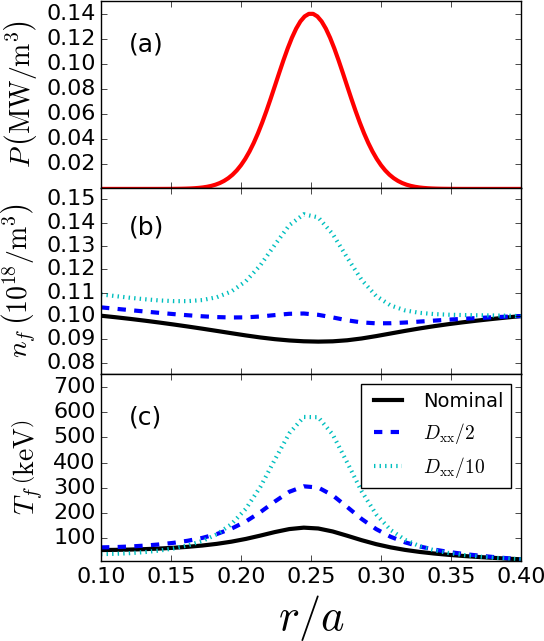}
   \caption{\label{radialplots} (a) Gaussian heat deposition profile
     applied in the \textsc{t3core} transport simulations, and the
     resulting (b) density, and (c) temperature profiles of the
     ICRH-heated $^3$He minority, the latter have the turbulent
  diffusivitives two scaled by the factors indicated. }
\end{figure}

Figure \ref{radialplots} shows a sample of our results, plotting the
radial profile of ICRH ions as a function of radius, showing the
effect of decreasing the diffusivities. Naturally, we define the
temperature for a non-Maxwellian distribution as
$T_f = \int \left(mv^2/3\right) F_0 \dv / n_f$. For sufficiently weak
transport (or sufficiently strong power injection), one obtains a
strong peak in temperature that depends quite strongly on the
turbulence.  When the turbulence is sufficiently suppressed so as to allow a sharp peak in
heated minority temperature, we observe the expected density pinch. In
the case of reduced turbulence, where the diffusion coefficients are
scaled down by factors of ten, the density of $^3$He is enhanced around the
heating location due to this ``thermodiffusion'' (a flux of particles against a temperature gradient). Eventually, particle
diffusion balances this effect to create the steady-state profiles
shown.

\begin{figure}
\centering
   \includegraphics[width=0.4\textwidth]{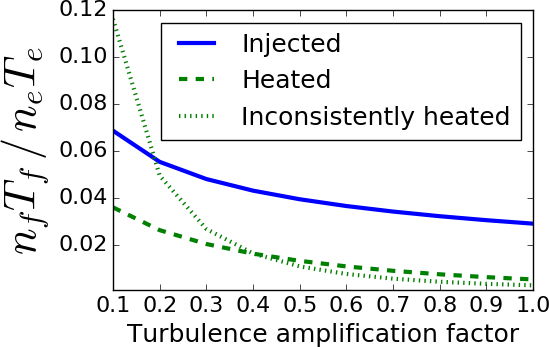}
   \caption{\label{pmax} Maximum pressure of fast ions (normalized to the electron pressure) as a function of the turbulence amplification factor. The ``inconsistently heated'' case is where a constant minority density is assumed throughout the domain when implementing the heating term. }
\end{figure}

We find that the NBI-like ``injected'' fast ions are less
sensitive to the turbulence than the ICRH ions: the peak density only
changes by a factor of two when the turbulence is weakened by the same
factor of ten, while the temperature changes by less than 15\% (300
keV for the nominal case). We demonstrate in Fig.~\ref{pmax} that the
peak pressure of ICRH-heated ions is more sensitive to the
turbulence than NBI-type ions and, by extension, alpha particles. The latter are capable of reaching higher pressures at a given amplitude of turbulence.

\begin{figure}
\centering
   \includegraphics[width=0.4\textwidth]{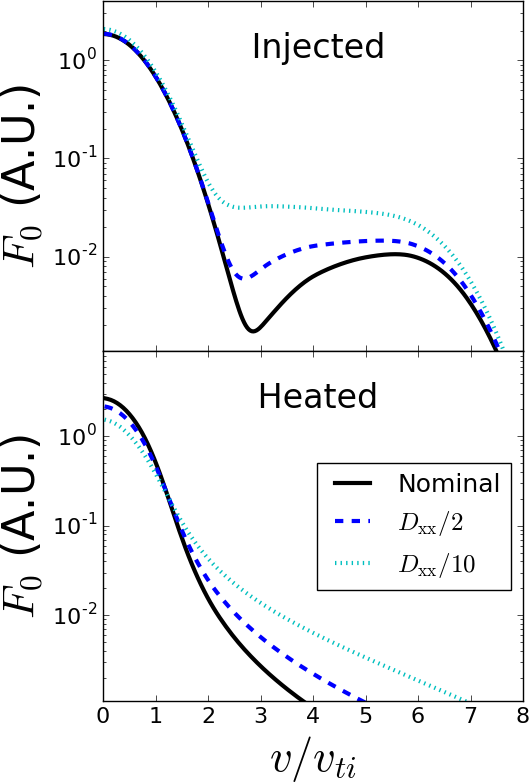}
   \caption{\label{f0} Modification of the distribution function of
     NBI-knock-ons (top) and ICRH-heated ions (bottom) at the peak
     heating radius $r= 0.25a $. To compare with the over-driven
     nominal case (solid black), the diffusion coefficients are scaled
     down by factors of two (dashed blue) and ten (dotted cyan) in
  each. }
\end{figure}

The microturbulence also has an effect on the velocity space
distribution (see Fig.~\ref{f0}), and there again we see ICRH-heated
fast ions being more sensitive. Note in Fig.~\ref{f0} that, even when
the turbulence is weak, it affects the velocity distribution of heated
ions at all energies, while the effect on the distribution of injected
ions is more localized around where $D_{rr}$ is dominant over
collisions. This localization in velocity space is seen even more
clearly in the distribution of alpha particles
\cite{wilkie_transport_2016}, which extends to yet higher energies,
where it is less affected by microturbulence.

\begin{figure}
\centering
   \includegraphics[width=0.5\textwidth]{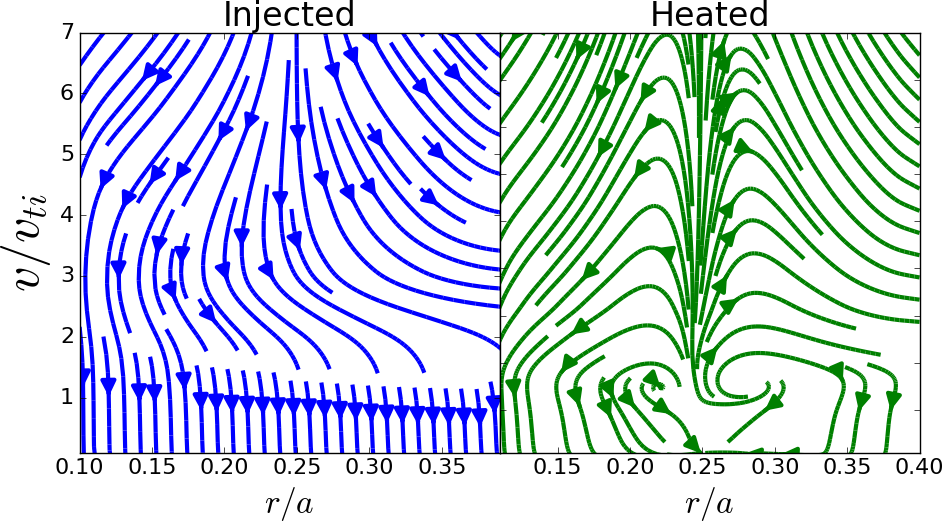}
   \caption{\label{streamplots} Stream plots showing trajectories of
     fast ions for the injected (left) and heated (right) case. The
     slope at each point is the normalized ratio
     $\Gamma_v a / \Gamma_r v_{ti}$. The case shown here is where the
     turbulence intensity is scaled down by a factor of ten. For reference, the injection energy of NBI ions is at $v_\mathrm{inj} \approx 6.3 v_{ti}$. }
\end{figure}

To understand the physical reason for this qualitative difference, in
Fig.~\ref{streamplots} examine the paths through phase space that each
type of fast ion takes. Recall that particles with moderately
suprathermal energies (where $v \sim 2 v_{ti}$) are most affected by the turbulence. Injected
ions slow down via collisions before they reach these energies, while
heated ions must ``pass through'' the turbulence-dominated part of
velocity space before becoming part of the distribution at high
energy. If the turbulence is too strong, heated particles are transported
radially before reaching high
energies. 

\section{Fast ion dilution model}

The trace approximation is needed when using the diffusive
form for the turbulent fluxes in Eqs.~\eqref{rflux} and
\eqref{vflux}. However, even when fast ions are not strictly trace,
they may still not participate directly in the drift wave dynamics. Instead, via
quasineutrality of the equilibrium, they take the place of the thermal
ions which are driving the turbulence. If this near-trace behavior can
be easily predicted, one could rescale the anomalous fluxes without
running additional turbulence simulations. This section seeks to
develop such a model for electrostatic turbulence.

When solving for the perturbed electrostatic potential $\phi$,
quasineutrality reads~\cite{abel_multiscale_2013}:
\begin{equation} \label{qneutrality} e \phi = -\, \frac{\sum_s Z_s \int
    \left\langle h_s \right\rangle_\mathbf{r} \dv}{\sum_s Z_s^2 \int
    \left( \partial F_{0s} / \partial E \right) \dv },
\end{equation}
where $h_s$ is the non-adiabatic part of the perturbed distribution of
species $s$, $F_{0s}$ is its equilibrium distribution function, and
$\left\langle \right\rangle_\mathbf{r}$ signifies the gyro-average at
fixed position in space. Suppose we have a simulation which contains
only singly-charged thermal ions and adiabatic electrons in
quasineutrality so that $ n_{i} = n_e$. In this case,
$e \phi \propto n_i / \left( n_i / T_i + n_e / T_e \right) = T_i/\left(1
  + \tau \right)$,
where $\tau \equiv T_i / T_e$. Now, introduce a population of adiabatic fast
ions with charge density $Z_f n_f$ such that now
$n_i + Z_f n_f = n_e$. Then, $e \phi \propto n_i / \left( n_i / T_i + n_e / T_e + Z_f^2 n_f / T_f \right) = T_i/\left(1 + \tau_\mathrm{eff} \right)$ , which serves to define an ``effective temperature ratio'' $\tau_\mathrm{eff}$, and $T_f$ is a suitable ``effective temperature'' of the fast ions, which may or may not be Maxwellian. In order for the computed $\phi$ to be the same
under these circumstances, we can account for the fast ions by
adjusting the temperature ratio of the original simulations
accordingly:
\begin{equation} \label{taueff}
  \tau_\mathrm{eff} = \left.  \left( \tau + \frac{T_i}{T_f} \frac{Z_f^2 n_f}{n_e} \right) \middle/ \left(1 - \frac{Z_f n_f}{n_e} \right) \right.,
\end{equation}
where $\tau$ is the original (physical) temperature ratio. This second term in the numerator can be neglected for sufficiently hot ions provided that $T_f \gg Z_f T_i$.  

The case used in this section is at $r=0.5a$ of ITER
scenario 10010100 from the ITPA database \cite{roach_2008_2008}, an
ELMy H-mode case of 50/50 deuterium/tritium. The physical parameters
at this flux surface are: $q=1.54$, $\hat{s} = 0.435$,
$\kappa = 1.49$, $a \kappa' = 0.38$, $\delta = 0.174$,
$a \delta' = 0.44$, and a Shafranov shift derivative of  $R_0'(r)= -0.084$. The bulk ions
are deuterium with a temperature $T_i = 0.86 T_e$, and nominally have
$a/L_{Ti} = 1.672$, while $a/L_{Te} = 1.732$ and
$a/L_{ne} = a/L_{ni} = 0$. The fast ions are considered to be
deuterium at a temperature of $T_f = 100 T_i$ and varying density,
with no density or temperature gradients. Due to the large number of
simulations required these simulations were of relatively low
resolution. The spatial grid is defined by $N_x = N_y = 24$,
$L_x \approx L_y = 10 \pi \rho_i$ perpendicular to the magnetic field,
and $N_\theta = 18$ along the magnetic field. Velocity space has 10
points in energy, and 26 pitch angles.

\begin{figure}
   \centering
   \includegraphics[width=0.7\textwidth]{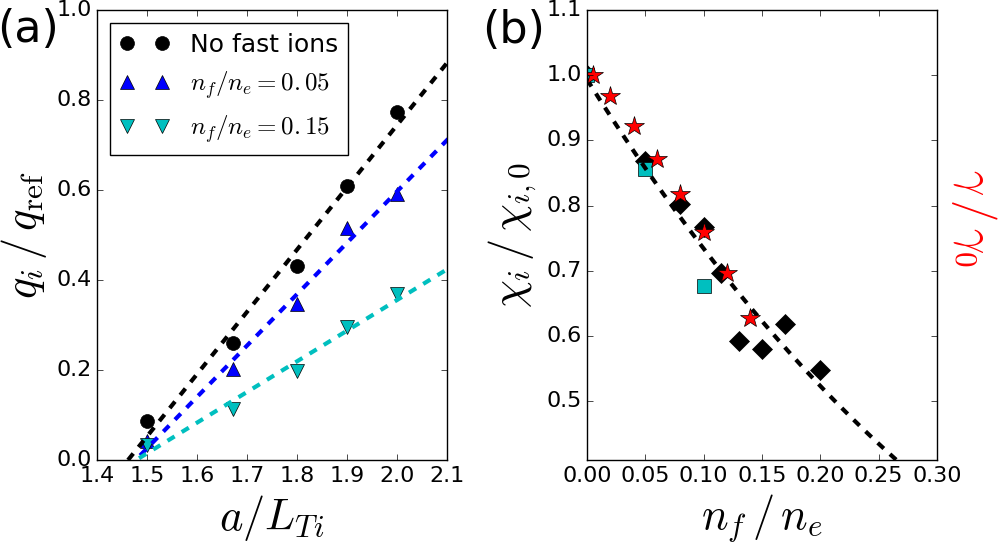}
   \caption{\label{dilution} (a) Turbulent deuterium heat flux,
     scanning in $a / L_{Ti}$ showing several fast ion
     concentrations. Dashed lines are the fitted diffusivities. (b)
     The thermal ion diffusivity as a function of fast ion
     concentration (black diamonds), a fitted (black dashed) line
     representing the presented model, and the dominant linear growth
     rate (red stars). Also shown are the results as determined from higher resolution nonlinear simulations (cyan squares).}
\end{figure}

Figure \ref{dilution}(a) shows the results of deuterium heat flux from
nonlinear \textsc{gs2} simulations near ITG marginality at several
fast ion concentrations. It is evident that dilution has little effect
on the critical gradient length scale:
$a/ L_{Ti0} \approx 1.49 \pm 0.04$, but it does reduce the
stiffness. The value of $\chi_i$ obtained for several fast ion
concentrations is shown in Fig.~\ref{dilution}(b). There, we also show
that $\chi_i$ scales similarly to the linear growth rate, as might be
expected from critical balance arguments \cite{barnes_critically_2011}. We choose a power law fit of these simulations to reflect
the behavior of Eq.~\eqref{taueff}, which results in the following
model:
\begin{equation} \label{dilutionmodel}
\chi_i \approx \chi_{i,0} \left(1 - \frac{Z_f n_f}{n_e} \right)^{2.9},
\end{equation}
where $\chi_{i,0}$ is the thermal ion diffusivity without any fast
ions. Note that the experimental scaling of thermal diffusivity with
respect to temperature ratio $\chi_i \propto \tau^{-3}$
\cite{petty_dependence_1999} is remarkably consistent with our results
when applying Eq.~\eqref{taueff}. We expect this model to be valid to
the extent that the following assumptions hold: the turbulence is
dominated by electrostatic dynamics; adiabatic electrons and fast ions
remain a good approximation; the temperature ratio continues to obey
the scaling of Ref.~\cite{petty_dependence_1999}; and one is in the
limit of high ``temperature'' fast ions (\emph{i.e.}~the fast ion
term vanishes in the denominator of Eq.~\eqref{qneutrality}). These
latter two assumptions can be relaxed by replacing the power in
Eq.~\eqref{dilutionmodel} and/or by including the additional term in
the numerator of Eq.~\eqref{taueff}. Furthermore, we note that
previous work done on fast ion dilution
\cite{tardini_thermal_2007,wilkie_microturbulent_2015} is also
consistent with this model.

By applying Eq.~\eqref{dilutionmodel}, a transport simulation of the
bulk plasma can easily and consistently respond to the stabilizing
presence of fast ions, at least to a leading approximation. This could
have implications for the formation of internal transport barriers,
and the intrepid modeller can optimise the heating profile to take
advantage of this effect.

\section*{Conclusion}

In this work, we explored the global non-Maxwellian transport of
isotropic heated minority ions, allowing their radial and velocity
distribution to arise naturally from the physics of collisions,
gyrokinetic microturbulence, and direct heating. Our aim was to
highlight the qualitative differences between their response to
microturbulence and that of fast ions such as those generated by NBI
or fusion reactions. Despite their very similar diffusion
coefficients, these differences fundamentally limit how robustly they
can mimic the turbulent transport of alpha particles, and this limitation is primarily due to how each type gains energy. However, their
relatively strong sensitivity could allow them to be a useful probe of
the turbulence, and as experimental validation for transport tools 
such as \textsc{t3core}. 

It has been experimentally observed that simultaneously applying NBI and IRCH heating simultaneously results in more heating than the sum of either of them alone \cite{krasilnikov_fundamental_2009}. Our results provide a possible explanation for this phenomenon. ICRH-heated minorities are more affected by the turbulence because they have the opportunity to transport radially before being heated. Therefore, if one heats ions via ICRH that are \emph{already} at high energy (via, for example, NBI), where turbulence has less effect, then one avoids the strong-transport region of phase space altogether.

Two important caveats to our results are the electrostatic and
isotropic assumptions. Electromagnetic fluctuations, even in primarily
electrostatic turbulence, could be important for trace transport at
sufficiently high plasma beta. Also, the fast ions considered here are
generally not isotropic in velocity space. In order for our results to
faithfully reflect the isotropic \emph{part} of such a velocity
distribution, we had to ignore the pitch-angle dependence of the
turbulent diffusion coefficients. We believe that our core result
(that ICRH ions are more sensitive to the microturbulence, having 
an impact on the distribution at all energies) is robust to relaxing
this assumption, but further study is expected and welcome.

The transport results presented here rely upon fast ions being trace,
which is a robust assumption for electrostatic turbulence at small
concentrations. Where fast ions might make up a more significant
fraction of the plasma, we presented a model to account for their
effect on the bulk plasma transport as a next approximation. Although the \textsc{t3core} simulations presented here do not employ this model, we
believe it will be very useful for future transport
simulations.

The authors would like to thank Yevgen Kazakov for the original
inspiration of high-energy heated minority and Torbj\"orn Hellsten for helpful discussion. Simulations were performed
on the SNIC cluster Hebbe (project nr.~SNIC2016-1-161) and the NERSC
supercomputer Edison. This work was supported by the Framework grant
for Strategic Energy Research (Dnr.~2014-5392) from
Vetenskapsr{\aa}det, the International Career Grant
(Dnr.~330-2014-6313), and the U.S. Department of Energy Office of
Fusion Energy Science (DEFG0293ER54197 and DEFC0208ER54964).

\appendix

\section{Onsager symmetry for quasilinear phase space transport}

The astute reader might notice a type of Onsager symmetry between
$D_{rv}$ and $D_{vr}$ at high energy in Fig.~\ref{diffcoeff}. A proof of Onsager symmetry for quasilinear radial transport of a Maxwellian species has already been shown \cite{sugama_transport_1996}, but this is not the case for nonlinear turbulence \cite{balescu_is_1991}.
In this appendix, we'll show that this symmetry holds for quasilinear transport even in $r$-$v$ phase space, as has been observed here and elsewhere \cite{waltz_quasilinear_2013}. This
is more than coincidence, and here we show that this is rigorous as
long as the magnetic drift velocity is dominant over the nonlinear
drift velocity in the gyrokinetic equation. 

The operator $\mathcal{L}$ is the left-hand side of the collisionless
gyrokinetic equation, so that when acting on the non-adiabatic
perturbed distribution $h$:
\begin{equation}\label{Ldef}
   \mathcal{L}\left[ h \right] \equiv \pd{h}{t} + \left(v_\| \mathbf{b} + \mathbf{v_d} + \mathbf{v}_\phi \right) \cdot \nabla h,
\end{equation}
where $\mathbf{v}_d$ is the magnetic drift velocity including the
$\nabla B$ and curvature drifts, and $\mathbf{v}_\phi \equiv \left(c/B
\right) \mathbf{b} \times \nabla \gyav{\phi}$. The unit vector
$\mathbf{b}$ points in the direction of the equilibrium magnetic
field.

What makes our transport calculations possible is that, for a trace
species only, $\mathcal{L}$ is a \emph{linear} operator, although it
does in general depend on $\phi$. Note that this is \emph{not}
equivalent to the quasilinear approximation used, for example, in
Refs.~\cite{angioni_gyrokinetic_2008} and
\cite{waltz_quasilinear_2013}, and is more generally applicable. The
quasilinear approximation is when one further assumes that the
$\mathbf{v}_\phi$ term in Eq.~\ref{Ldef} is negligible. This becomes
more accurate at high energy, when $\mathbf{v}_d \gg \mathbf{v}_\phi$.

Now, consider the following expressions for the diffusion coefficients:
\begin{equation} \label{Drrdef}
   D_{r r}  = \turbav{\sum_{\sigma_\|} \int\mathcal{L}^{-1} \left[\mathbf{v}_\phi \cdot \nabla r  \right] \left( \mathbf{v}_\phi  \cdot \nabla r \right)\jacobv \mathrm{d}\lambda}
\end{equation} \begin{equation} \label{Drvdef}
D_{r v}  = \turbav{\sum_{\sigma_\|} \int\mathcal{L}^{-1} \left[ \pd{\gyav{\phi}}{t}  \right] \frac{Z e}{mv} \left( \mathbf{v}_\phi \cdot \nabla r \right)\jacobv \mathrm{d}\lambda}
\end{equation}
\begin{equation} \label{Dvrdef}
D_{v r}  = \turbav{\sum_{\sigma_\|} \int\mathcal{L}^{-1} \left[ \mathbf{v}_\phi \cdot \nabla r \right] \frac{Z e}{mv} \left( \pd{\gyav{\phi}}{t} \right) \jacobv \mathrm{d}\lambda}
\end{equation}
\begin{equation} \label{Dvvdef}
D_{v v}  = \turbav{\sum_{\sigma_\|} \int\mathcal{L}^{-1} \left[ \pd{\gyav{\phi}}{t} \right] \frac{Z^2 e^2}{m^2v^2} \left( \pd{\gyav{\phi}}{t} \right) \jacobv \mathrm{d}\lambda}.
\end{equation}
Here, $\jacobv$ is the velocity space Jacobian for coordinates $v$ and
$\lambda = v_\perp^2 / v^2$, $\sigma_\|$ is the sign of the parallel
velocity (which would otherwise ambiguous in these coordinates), and
$\turbav{ }$ represents a time-average and a spatial average over the
flux-tube. Consider the off-diagonal terms defined in
Eqs.~\ref{Drvdef} and \ref{Dvrdef}. If we assume that $\mathcal{L}$
does not depend on $\phi$ (thus has no explicit time dependence), then
we can show that $D_{rv} = D_{vr}$ by Fourier transforming in
time. For convenience, let $\mathcal{D} \equiv \mathcal{L} - \partial/ \partial t $. Recalling the definition of time average to be an
integral over time:
\begin{equation}
   \int  \mathcal{L}^{-1}\left[ \mathbf{v}_\phi(t) \cdot \nabla r \right] \pd{\gyav{\phi}}{t} \,dt 
\end{equation}
\begin{equation}
   = \int  \pd{\gyav{\phi}}{t} \int  \,\tilde{\mathcal{L}}^{-1}\left[ \mathbf{v}_\phi(t) \cdot \nabla r \right]\left( \omega \right) e^{i\omega t} d\omega\,\,dt 
\end{equation}
\begin{equation}
   = \int  \pd{\gyav{\phi}}{t} \int  \,\frac{\tilde{\mathbf{v}}_\phi(\omega)\cdot \nabla r }{-i \omega + \mathcal{D} }  e^{i\omega t} \,d\omega\,dt  
\end{equation}
\begin{equation}
   = \int  \int \int \pd{\gyav{\phi}}{t}  \,\frac{\mathbf{v}_\phi(t')\cdot \nabla r }{-i \omega + \mathcal{D} }  e^{i\omega \left(t-t'\right)} \,dt'\,d\omega\,dt  
\end{equation}
\begin{equation}
   = \int \int \int  \mathbf{v}_\phi(t)\cdot \nabla r \frac{\partial \gyav{\phi} / \partial t'}{-i \omega + \mathcal{D} }  e^{i\omega \left(t'-t\right)} \,dt'\,d\omega\,dt  
\end{equation}
\begin{equation}
   = \int \mathcal{L}^{-1}\left[ \pd{\gyav{\phi}}{t'} \right] \mathbf{v}_\phi(t') \cdot \nabla r \,dt' ,
\end{equation}
which shows that Eqs.~\ref{Drvdef} and \ref{Dvrdef} are equivalent. The
key approximation was that $\mathcal{D}$ does not depend on time,
which is the case for the quasilinear approximation. Therefore,
while not generally applicable in nonlinear turbulence, 
Onsager symmetry is expected at high energy to the extent that
$\mathbf{v}_\phi \ll \mathbf{v}_d$.
\section*{References}

\bibliographystyle{unsrt}
\bibliography{zotero}

\begin{thebibliography}{10}

\bibitem{kazakov_resonant_2015}
Ye.O. Kazakov, D.~Van~Eester, R.~Dumont, and J.~Ongena.
\newblock On resonant {ICRF} absorption in three-ion component plasmas: a new
  promising tool for fast ion generation.
\newblock {\em Nuclear Fusion}, 55(3):032001, March 2015.

\bibitem{pusztai_turbulent_2016}
Istv{\'a}n Pusztai, George~J Wilkie, Yevgen~O Kazakov, and T{\"u}nde
  F{\"u}l{\"o}p.
\newblock Turbulent transport of {MeV} range cyclotron heated minorities as
  compared to alpha particles.
\newblock {\em Plasma Physics and Controlled Fusion}, 58(10):105001, November
  2016.

\bibitem{angioni_gyrokinetic_2008}
C.~Angioni and A.~G. Peeters.
\newblock Gyrokinetic calculations of diffusive and convective transport of
  $\alpha$ particles with a slowing-down distribution function.
\newblock {\em Physics of Plasmas}, 15(5):052307, 2008.

\bibitem{wilkie_microturbulent_2015}
G.~J. Wilkie.
\newblock {\em Microturbulent transport of non-{Maxwellian} alpha particles}.
\newblock {PhD} thesis, University of Maryland, 2015.

\bibitem{waltz_quasilinear_2013}
R.~E. Waltz, E.~M. Bass, and G.~M. Staebler.
\newblock Quasilinear model for energetic particle diffusion in radial and
  velocity space.
\newblock {\em Physics of Plasmas}, 20(4):042510, 2013.

\bibitem{hauff_electrostatic_2009}
T.~Hauff, M.~J. Pueschel, T.~Dannert, and F.~Jenko.
\newblock Electrostatic and magnetic transport of energetic ions in turbulent
  plasmas.
\newblock {\em Physical Review Letters}, 102(7), February 2009.

\bibitem{pueschel_anomalous_2012}
M.J. Pueschel, F.~Jenko, M.~Schneller, T.~Hauff, S.~G{\"u}nter, and G.~Tardini.
\newblock Anomalous diffusion of energetic particles: connecting experiment and
  simulations.
\newblock {\em Nuclear Fusion}, 52(10):103018, October 2012.

\bibitem{wilkie_validating_2015}
G.~J. Wilkie, I.~G. Abel, E.~G. Highcock, and W.~Dorland.
\newblock Validating modeling assumptions of alpha particles in electrostatic
  turbulence.
\newblock {\em Journal of Plasma Physics}, 81(03):905810306, 2015.

\bibitem{helander_collisional_2002}
P.~Helander and D.~Sigmar.
\newblock {\em Collisional {Transport} in {Magnetized} {Plasmas}}.
\newblock Cambridge University Press, Cambridge, 2002.

\bibitem{stix_fast-wave_1975}
Thomas~Howard Stix.
\newblock Fast-wave heating of a two-component plasma.
\newblock {\em Nuclear Fusion}, 15(5):737, 1975.

\bibitem{wilkie_transport_2016}
G.~J. Wilkie, I.~G. Abel, M.~Landreman, and W.~Dorland.
\newblock Transport and deceleration of fusion products in microturbulence.
\newblock {\em Physics of Plasmas}, 23(6):060703, June 2016.

\bibitem{dorland_electron_2000}
W.~Dorland, F.~Jenko, M.~Kotschenreuther, and B.~N. Rogers.
\newblock Electron {Temperature} {Gradient} {Turbulence}.
\newblock {\em Physical Review Letters}, 85(26):5597, 2000.

\bibitem{roach_2008_2008}
C.M. Roach, M.~Walters, R.V. Budny, F.~Imbeaux, T.W. Fredian, M.~Greenwald,
  J.A. Stillerman, D.A. Alexander, J.~Carlsson, J.R. Cary, F.~Ryter, J.~Stober,
  P.~Gohil, C.~Greenfield, M.~Murakami, G.~Bracco, B.~Esposito, M.~Romanelli,
  V.~Parail, P.~Stubberfield, I.~Voitsekhovitch, C.~Brickley, A.R. Field,
  Y.~Sakamoto, T.~Fujita, T.~Fukuda, N.~Hayashi, G.M.D Hogeweij,
  A.~Chudnovskiy, N.A. Kinerva, C.E. Kessel, T.~Aniel, G.T. Hoang, J.~Ongena,
  E.J. Doyle, W.A. Houlberg, A.R. Polevoi, {ITPA Confinement Database and
  Modelling Topical Group}, and {ITPA Transport Physics Topical Group}.
\newblock The 2008 {Public} {Release} of the {International} {Multi}-tokamak
  {Confinement} {Profile} {Database}.
\newblock {\em Nuclear Fusion}, 48(12):125001, December 2008.

\bibitem{abel_multiscale_2013}
I~G Abel, G~G Plunk, E~Wang, M~Barnes, S~C Cowley, W~Dorland, and A~A
  Schekochihin.
\newblock Multiscale gyrokinetics for rotating tokamak plasmas: fluctuations,
  transport and energy flows.
\newblock {\em Reports on Progress in Physics}, 76(11):116201, November 2013.

\bibitem{barnes_critically_2011}
M.~Barnes, F.~I. Parra, and A.~A. Schekochihin.
\newblock Critically {Balanced} {Ion} {Temperature} {Gradient} {Turbulence} in
  {Fusion} {Plasmas}.
\newblock {\em Physical Review Letters}, 107(11), September 2011.

\bibitem{petty_dependence_1999}
C.~C. Petty, M.~R. Wade, J.~E. Kinsey, R.~J. Groebner, T.~C. Luce, and G.~M.
  Staebler.
\newblock Dependence of heat and particle transport on the ratio of the ion and
  electron temperatures.
\newblock {\em Physical review letters}, 83(18):3661, 1999.

\bibitem{tardini_thermal_2007}
G~Tardini, J~Hobirk, V.G Igochine, C.F Maggi, P~Martin, D~McCune, A.G Peeters,
  A.C.C Sips, A~St{\"a}bler, J~Stober, and the ASDEX~Upgrade Team.
\newblock Thermal ions dilution and {ITG} suppression in {ASDEX} {Upgrade} ion
  {ITBs}.
\newblock {\em Nuclear Fusion}, 47(4):280--287, April 2007.

\bibitem{krasilnikov_fundamental_2009}
A~V Krasilnikov, D~Van~Eester, E~Lerche, J~Ongena, V~N Amosov, T~Biewer,
  G~Bonheure, K~Crombe, G~Ericsson, B~Esposito, L~Giacomelli, C~Hellesen,
  A~Hjalmarsson, S~Jachmich, J~Kallne, Yu~A Kaschuck, V~Kiptily, H~Leggate,
  J~Mailloux, D~Marocco, M-L Mayoral, S~Popovichev, M~Riva, M~Santala, M~Stamp,
  V~Vdovin, A~Walden, and {JET EFDA Task Force Heating and JET EFDA
  contributors}.
\newblock Fundamental ion cyclotron resonance heating of {JET} deuterium
  plasmas.
\newblock {\em Plasma Physics and Controlled Fusion}, 51(4):044005, April 2009.

\bibitem{sugama_transport_1996}
H.~Sugama, M.~Okamoto, W.~Horton, and M.~Wakatani.
\newblock Transport processes and entropy production in toroidal plasmas with
  gyrokinetic electromagnetic turbulence.
\newblock {\em Physics of Plasmas}, 3(6):2379, 1996.

\bibitem{balescu_is_1991}
R.~Balescu.
\newblock Is {Onsager} symmetry relevant in the transport equations for
  magnetically confined plasmas?
\newblock {\em Physics of Fluids B: Plasma Physics}, 3(3):564, 1991.

\end{thebibliography}

\end{document}